\documentclass[pra,twocolumn,showpacs,superscriptaddress,
groupeaddress, preprintnumbers,amsmath,amssymb,tightenlines]{revtex4}
\usepackage{graphicx}
\newcommand{\beq}{\begin{equation}}
\newcommand{\eeq}{\end{equation}}
\newcommand{\bea}{\begin{eqnarray}}
\newcommand{\eea}{\end{eqnarray}}
\newcommand{\ba}{\begin{array}}
\newcommand{\ea}{\end{array}}
\newcommand{\bc}{\begin{center}}
\newcommand{\ec}{\end{center}}
\newcommand{\lsimeq}{\alt}
\newcommand{\gsimeq}{\agt}

\newcommand{\bml}{\begin{subequations}}
\newcommand{\eml}{\end{subequations}}
\newcommand{\commentout}[1]{{}}
\newcommand{\bk}{{\bf k}}

\newcommand{\br}{{\bf r}}

\newcommand{\K}{{\cal K}}

\newcommand{\adag}{a^\dagger}
\newcommand{\bdag}{b^\dagger}
\newcommand{\cdag}{c^\dagger}
\newcommand{\psidag}{\psi^\dagger}

\newcommand{\alphadag}{\alpha^\dagger}
\newcommand{\ddt}{\frac{d}{dt}}
\newcommand{\half}{\hbox{$\frac{1}{2}$}}

\newcommand{\HC}{{\rm H.c.}}
\newcommand{\eq}[1]{(\ref{#1})}
\newcommand{\etal} {{\it et al.\/}}
\newcommand{\ibid} {{\it ibid. \/}}
\newcommand{\vol}[1]{{\bf #1}}
\newcommand{\comment}[1]{{}}

\newcommand{\rbpot}{$^{87}\rm{Rb}\,$-$^{40}$K }

\begin{document}
\title{New Directions in Degenerate Dipolar Molecules via
Collective Association}
\author{Matt Mackie}
\affiliation{QUANTOP--Danish National Research Foundation
Center for Quantum Optics, Department of Physics and Astronomy,
University of Aarhus, DK-8000 Aarhus C, Denmark}
\author{Olavi Dannenberg}
\affiliation{Helsinki Institute of Physics, PL 64, FIN-00014
Helsingin yliopisto, Finland}
\author{Jyrki Piilo}
\affiliation{School of Pure and Applied Physics,
University of KwaZulu-Natal, Durban 4041, South Africa}
\author{Kalle-Antti Suominen}
\affiliation{Department of Physics, University of Turku, FIN-20014
Turun yliopisto, Finland}
\affiliation{Helsinki Institute of Physics, PL 64, FIN-00014
Helsingin yliopisto, Finland}
\author{Juha Javanainen}
\affiliation{Department of Physics, University of Connecticut,
Storrs, Connecticut, 06269-3046}
\date{\today}

\begin{abstract}
We survey results on the creation of heteronuclear Fermi molecules by
tuning a degenerate Bose-Fermi mixture into the neighborhood of an
association resonance, either photoassociation or Feshbach, as well as the
subsequent prospects for Cooper-like pairing between atoms and molecules.
In the simplest case of only one molecular state, corresponding to either
a Feshbach resonance or one-color photoassociation, the system displays
Rabi oscillations and rapid adiabatic passage between a Bose-Fermi
mixture of atoms and fermionic molecules. For two-color photoassociation,
the system admits stimulated Raman adiabatic passage (STIRAP) from a
Bose-Fermi mixture of atoms to stable Fermi molecules, even in the
presence of particle-particle interactions. By tailoring the STIRAP
sequence it is possible to deliberately convert only a fraction of the
initial atoms, leaving a finite fraction of bosons behind to induce
atom-molecule Cooper pairing via density fluctuations; unfortunately, this
enhancement is insufficient to achieve a superfluid transition
with present ultracold technology. We therefore propose the use of an
association resonance that converts atoms and diatomic molecules (dimers)
into triatomic molecules (trimers), which leads to a crossover from a
Bose-Einstein condensate of trimers to atom-dimer Cooper pairs. Because
heteronuclear dimers may possess a permanent electric dipole moment,
this overall system presents an opportunity to investigate novel
microscopic physics.
\end{abstract}


\maketitle

\section{Introduction}

It has recently come to the fore that
collective association can actually produce a quantum degenerate
molecules, serving as a compliment to the development of buffer-gas
cooling~\cite{WEI98} and Stark-deceleration~\cite{BET00} techniques.
The trail to molecular Bose-Einstein condensate (BEC) was
pioneered by experiments~\cite{WYN00} with photoassociation of $^{87}$Rb
condensate that were just on the verge~\cite{KOS00} of coherent
atom-molecule conversion. Next-generation experiments were aimed at the
strongly interacting regime~\cite{MCK02,PRO03}, and probed the
predicted~\cite{BOH99,KOS00,JAV02} photoassociation rate limit. As of
late, laser experiments~\cite{THE04} have
confirmed~\cite{FED96,BOH97,KOS00} an optically tunable BEC scattering
length, no small feat in the face of rapid spontaneous decay.
Meanwhile, groundbreaking magnetoassociation experiments
demonstrated a magnetically tunable scattering
length~\cite{COU98,INO98}, and led to the controlled collapse of a
condensate with atom bursts emanating from remnant
condensate~\cite{DON01}. Subsequent coherent oscillations between burst
and remnant~\cite{DON02} were interpreted as evidence for atom-molecule
coherence~\cite{KOK02,MAC02a,KOE03a,DUI03}. In the search for the
neutral-atom analogue of superconducting Cooper pairs of
electrons~\cite{REG04,ZWI04}, ultracold~\cite{REG03,STR03,CUB03} and
condensate~\cite{JOC03,GRE03,ZWI03} molecules were unambiguously created
via magnetoassociation of fermionic atoms~\cite{CAR03,FAL04,JAV04}.
Similarly, quantum degenerate molecules have been observed and
characterized in magnetoassociation of atomic BEC~\cite{KXU03}.

Collective association is a general theory to describe a
quantum-degenerate gas tuned near a photoassociation or Feshbach
resonance. The former is a resonance where two colliding atoms absorb a
photon and jump to a bound molecular state~\cite{THO87}, whereas the
latter is a resonance where one atom from a colliding pair spin flips in
the presence of a magnetic field~\cite{STW76,TIE92}, known affectionately
as magnetoassociation. Some of the first theoretical investigations into
association of a quantum degenerate gas pointed to strongly enhanced
molecule formation~\cite{BUR97,JAV98}, and a focus on Bose-Einstein
condensates~\cite{DRU98,JAV99,VAR01,TOM98,VAB99,YUR99,MIE00} in
particular led to the idea of Bose-enhanced molecule
production~\cite{MAC00}, or so-called superchemistry~\cite{HEI00}.
However, it soon became clear that this idea applied equally well to
degenerate Fermi and Bose-Fermi mixtures, indicating the possibility for
changes in statistics, i.e., creating Bose molecules from Fermi
mixtures~\cite{TIM01,HOL01,OHA02,SEA02,MAC04} or Fermi molecules from
Bose-Fermi mixtures~\cite{DAN03,WOU03}, and thereby taking superchemistry
to a new level. Moreover, collective association~\cite{SHA99} of
Bose-Fermi mixtures~\cite{TRU01} in particular offers a convenient source
of ultracold polar molecules, which can be used for quantum
computation~\cite{DEM02} and searches for the standard-model-violating
electric dipole moment of the electron~\cite{KOZ95}.

So motivated, our attention has therefore turned to
resonant association in Bose-Fermi mixtures of
atoms, and the subsequent possibility of atom-molecule
Cooper pairing. While the Raman formation of ground state Fermi molecules
appears feasible in practice~\cite{MAC04}, the subsequent off-resonant
transition to a superfluid of atom-molecule Cooper pairs occurs out of
reach of present ultracold technology~\cite{MAC04}; hence, an association
resonance for {\em trimers} is apparently necessary to experimentally
realize atom-dimer Cooper pairing~\cite{MAC04b}. The purpose of this
article is to survey our results on collective Fermi molecule formation,
in order to highlight future research avenues in the physics of quantum
degenerate dipolar gases.

The survey herein is outlined as follows.
In Sec.~\ref{SHORT}, we discuss the broad features of converting a
degenerate Bose-Fermi mixture into Fermi molecules via a collective
association resonance, in particular Rabi-like oscillation and rapid
adiabatic passage. In Sec.~\ref{RAM_COOP}, the focus is on the formation
of stable Fermi molecules with Raman photoassociation of a Bose-Fermi
mixture, and the subsequent possibility for boson-induced Cooper pairing
between atoms and molecules, where Raman photoassociation converts a
fraction of the initial mixture into molecules and the leftover bosons
enhance the atom-molecule interaction. Since this enhancement is
insufficient to induce pairing within reach of present ultracold
technology, Sec.~\ref{TRIMERS} explores the possibility of driving
atom-molecule Cooper pairing with an association resonance that converts
atom-dimer pairs into trimers, i.e., a crossover from a BEC of trimers to
atom-dimer Cooper pairs. Besides a summary, Sec.~\ref{OUT} identifies
some open questions regarding dipolar particle interactions.

\section{Shortcut to Degenerate Fermi Molecules via Collective
Association}
\label{SHORT}

Analogous with coherent optical transients in few level atomic
systems~\cite{ALL87}, photoassociation of a BEC has been predicted to
induce Rabi-like oscillations between atomic and molecular
condensates~\cite{JAV99,HEI00,VAR01}, whereby an entire gas of, say,
two million Bose-condensed atoms are collectively converted into a million
molecules that are, in turn, collectively converted back into (roughly)
two million atoms, {\it ad infinitum}. Another interesting possibility
arises because the ground state of the system is all atoms for large
positive detunings (far below threshold) and all molecules for large
negative detunings (far above threshold), so that a slow sweep of the
laser detuning from one extreme to the other will collectively convert a
BEC of atoms into a BEC of molecules~\cite{JAV99}. Incidentally, it was a
particular combination of these two concepts, applied instead to
magnetoassociation, that led to the observation\cite{DON02} of collective
Ramsey fringes between an atomic condensate and a small fraction of
molecular condensate dressed by dissociated atom
pairs~\cite{KOK02,MAC02a,KOE03a,DUI03}. Given a degenerate mixture of Bose
and Fermi gases~\cite{TRU01}, will similar coherent phenomena
manifest in free-bound association degenerate Fermi
molecules?

To address this question, we model
an ideal degenerate Bose-Fermi mixture of atoms
coupled by either a Feshbach or photoassociation resonance to a
Fermi-degenerate gas of molecules.
But first we make some simplifications to allow for ease of modeling.
First, by making a proper~\cite{DAN03} unitary transformation, any
explicit chemical potential can be absorbed into the detuning and
forgotten. Second, collective association occurs on a timescale much
faster than trapped-atom motion, allowing neglect of the kinetic
energies, and justifying omission of an explicit trap. Third, any Fermi
energies lie within the Wigner
threshold regime, so that the coupling
$\K$ can be taken as the same for all modes. Finally, we work in
momentum space, but retain only the $\bk=0$ condensate modes since
Bose stimulation favors these transitions over $\bk\neq 0$
modes. The simplified Hamiltonian reads
\beq
H=\sum_\bk\left[\delta\bdag_\bk b_\bk
  -\half\kappa (\bdag_\bk a_\bk c +\cdag\adag_\bk b_\bk)\right],
\label{SIMP_HAM}
\eeq
where $c$ annihilates a condensate atom, $a_\bk$ ($b_\bk$)
annihilates an atom (molecule) with wavevector $\bk$, and $\kappa={\cal
K}/\sqrt{V}$.

\begin{figure}[b]
\centering
\includegraphics[width=8.0cm]{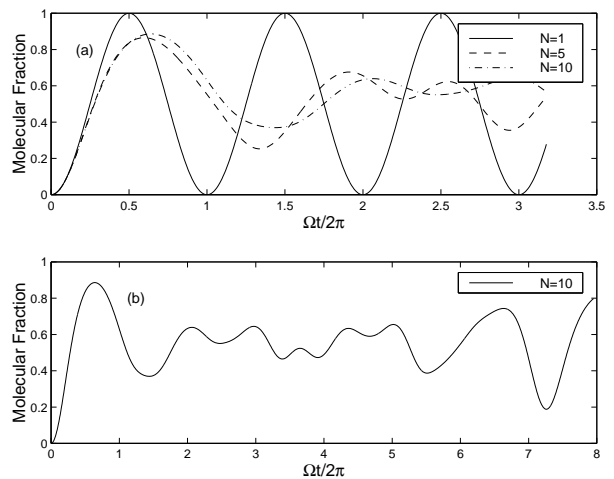}
\caption{Rabi-like oscillations in the fraction of a Fermi-Bose gas of
atoms converted to Fermi molecules.
The initial number of bosonic and fermionic atoms are equal, i.e.,
$N_B=N_F=N$. (a) The oscillations are complete for an initial
number of atoms
$N=1$, while for
$N=5$ many-body effects lead to frequency-shifted oscillations that are
incomplete and collapse. Better short-time agreement with the $N=1$ result
is obtained for $N=10$. (b) The oscillations eventually
revive.}
\label{RABI}
\end{figure}
\vspace{-0.5cm}

From the
Heisenberg equations of motion for an on-resonance system
($\delta=0$)~\cite{DAN03}, it is easy to show that for
equal numbers of bosonic and fermionic atoms, $N_B=N_F=N$, the system that
evolves with the characteristic frequency $\Omega=\sqrt{N}\kappa$.
Using Fock states, said intuition is confirmed by solving the
Schr\"odinger equation numerically, the results of which
are shown in Fig.~\ref{RABI}. For the two-body case
($N=1$), a complete oscillation between Fermi-Bose atoms and Fermi
molecules occurs in a time $2\pi/\Omega$. However, for $N=5$ quantum
many-body fluctuations not only frustrate complete conversion, but also
shift the oscillation frequency and lead to collapse and revival.
Increasing the initial particle number to $N=10$ gives better short-time
agreement with the two-body result. This behavior is exactly analogous to
the single-component bosonic case~\cite{JAV99,VAR01}. Although limited
computational resources preclude explicit investigation, based on the
bosonic analogy~\cite{JAV99,VAR01} we fully expect the first half
oscillation to be complete for large particle number.

\begin{figure}
\centering
\includegraphics[width=8.0cm]{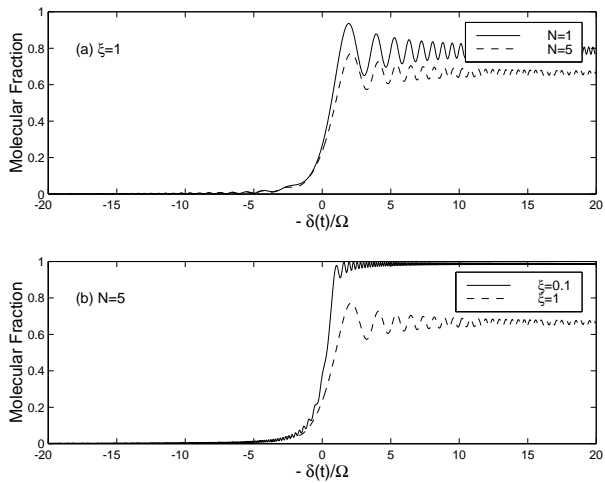}
\caption{Rapid adiabatic passage from a Fermi-Bose gas of atoms to
Fermi molecules. The detuning is swept
as $\delta(t)=-\xi\Omega^2t$, and $N_B=N_F=N$. (a) For borderline
adiabaticity, 
$\xi=1$, increasing the
number of initial atoms from $N=1$ to $N=5$ indicates that many-body
effects reduce the efficiency. (b) For $N=5$ and $\xi=0.1$,
near-unit conversion is still possible, despite many-body effects.}
\label{SWEEP}
\end{figure}

From the Hamiltonian~\eq{SIMP_HAM}, it should also be clear that the
system will favor all atoms for large positive detunings, while favoring
all molecules for large negative detunings. With $\Omega$ established as
the characteristic frequency for collective atom-molecule conversion,
changes in the detuning that are of the order of
$\Omega$, and occur over a time of order $\Omega^{-1}$, should
qualify as adiabatic. Hence, if the detuning is swept in a
linear fashion according to $\delta=-\xi\Omega^2 t$, then dimensionless
sweep rates $\xi\lsimeq 1$ should enable rapid adiabatic passage to
degenerate Fermi molecules.
Our suspicions are again
corroborated by a Fock-state-based numerical solution to the Schr\"odinger
equation, shown in Fig.~\ref{SWEEP}. While many-body effects appear to
rather seriously affect the efficiency of a marginally adiabatic sweep
($\xi=1$) compared to the $N=1$ case, the difference between $N=5$ and
$N=10$ (not shown) is in fact small. Overall, many-body effects are
expected to be weak for near-adiabatic sweeps ($\xi\sim 1$), and
vanishing for sweeps that are deep-adiabatic ($\xi\lsimeq 0.1$), again in
agreement with our BEC experience~\cite{JAV99}. 

For an estimate of explicit numbers we eschew
photoassociation because of the losses associated with the
electronically-excited state, and focus on the atom-molecule
coupling provided by the  Feshbach resonance located at
$B_0=534\,$G~\cite{SIM03}, which has a width $\Delta_R=4\,$G and an
associated zero-field Fermi-Bose-atom scattering length
$a_{aa}^{FB}=-17.8\,$nm. Accordingly, the atom-molecule coupling is
${\cal K}\approx (4\pi|a_{aa}^{FB}|
  \mu_{\rm Bohr}\Delta_R/m_r)^{1/2}
    =0.14\,\rm{cm}^{3/2}\,\rm{s}^{-1}$,
where we have estimated the difference between the Fermi-Bose atom pair
and molecular magnetic moments to be equal to the Bohr magneton
$\mu_{\rm Bohr}$.
Assuming $N_B=10^5$ condensate atoms in a trap with respective radial and
axial frequencies $\omega_r/2\pi=215\,$Hz and
$\omega_a/2\pi=16.3\,$Hz~\cite{MOD02}, the density of bosons is
$\rho_B=8.1\times 10^{13}\,\rm{cm}^{-3}$. As for the fermions, we assume
a modest number, say, $N_F=10^3$, which has three consequences: (i)  the
atomic BEC will act as a reservoir, thus absorbing any heat created by
holes in the Fermi sea~\cite{TIM01b}; (ii) barring an unfortunately large
scattering length for Bose-atom and Fermi-molecule collisions, we can
neglect the possibility of any Fermi-Bose collapse
instabilities~\cite{MOD02}; (iii) the size of the Fermi cloud
($R_F=8.3\,\mu\rm{m}$) is smaller than the BEC ($R_B=10\,\mu\rm{m}$), so
that overlap is not an issue. Moreover, for
$N_B\gg N_F$, the timescale for atom-molecule conversion is
$\tau_{a2m}\sim (\sqrt{\rho_B}\,\K)^{-1}=8.2\times 10^{-7}\,\rm{s}$.
This timescale is safely below the fastest timescale for trapped-atom
motion
$\tau_t=(\omega_r/2\pi)^{-1}=4.7\times 10^{-3}\,\rm{s}$,
justifying our neglect of trap dynamics and the kinetic
energy; physically put, this means that the Fermi energy is negligible
compared to the atom-molecule coupling strength.

It is important to mention two neglected complications: noncondensate
modes and interactions between particles. On the matter of dissociation to
noncondensate modes, and the related pair
correlations\cite{KOK02,MAC02a,KOE03a,DUI03}, we note that in the
all-boson case~\cite{YUR03} such transitions can be neglected for a sweep
directed as in Fig.~\ref{SWEEP}, i.e., for
$\dot\delta<0$. Given the success of the analogy so far, similar is
expected for the Fermi molecule system. As for collisions between
particles, they are described in terms of the coupling strength is
$\lambda=2\pi\hbar a/m^*$, where $a$ is the $s$-wave scattering length,
$m^*$ is the mass of the atom or the reduced mass of the atom-atom
(atom-molecule) pair. An estimate for $\lambda$ indicates that it is
negligible compared to the atom-molecule coupling for a typical \rbpot
system and, besides, particle interactions have little effect on sweeps
across resonance in the boson case~\cite{ISH04} and similar is expected
for Fermi molecules.

\section{Raman Photoassociation of Stable Fermi Molecules and their
Subsequent Cooper Pairing with Atoms}
\label{RAM_COOP}

Given that a degenerate mixture of
Bose and Fermi atoms admits basic collective association, the natural
question to ask is whether Raman photoassociation to the ground molecular
state will be efficient, since for bosons it is not a trivial matter to
overcome the condensate mean-field shifts. If so, the follow up to this
question is whether it
possible to realize Cooper pairing, not between atoms or molecules,
but instead between atoms and molecules.

First introduced to explain
superconductivity, anomalous quantum correlations between
two degenerate electrons with equal and opposite momenta--Cooper
pairs--are due physically to an electron-electron attraction
mediated by the exchange of lattice-vibration-generated
phonons~\cite{BAR57}, and are a manifestation of fermionic
superfluidity~\cite{TIN75}. Anomalous pairing between different chemical
species was immediately suggested to explain the larger excitation energy
for nuclei with even rather than odd numbers of
nucleons~\cite{BOH58}, although it turned out that interspecies
pairing plays the dominant role.
Today quantum matter optics offers a means to explore condensed-matter
and nuclear physics by proxy, such as the pairing of fermions in atomic
traps and nuclei~\cite{HEI03}. Here we point out that, while the
dipolar-dipolar case has been considered, Cooper pairing between
non-dipolar atoms and dipolar molecules offers a novel avenue of
investigation.

In this section we survey Raman
photoassociation~\cite{VAR97,MAC00,HOP01,DAM03} of Bose-Fermi mixtures of
atoms~\cite{TRU01}, and the subsequent prospects for Cooper pairing
between different chemical species (i.e., atoms and molecules). First, we
demonstrate that an arbitrary number of stable Fermi molecules can be
created via fractional~\cite{MAR91} stimulated Raman adiabatic passage
(STIRAP~\cite{BER98}), which is feasible because, contrary to bosonic
systems~\cite{HOP01}, collisional interactions can be negligible. Density
fluctuations in the condensate leftover from the photoassociation process
then replace the vibrating ion lattice of the
superconductor~\cite{HEI00b}, and the subsequent phonon exchange can
enhance the intrafermion attraction. We find that a typical attraction is
enhanced, but this enhancement is insufficient for a transition to
atom-molecule Cooper pairs within reach of present ultracold technology.

We model
a Bose-Fermi mixture of atoms
coupled by heteronuclear photoassociation to electronically-excited
Fermi molecules, which is favored over homonuclear transitions
for well resolved resonances. The excited molecules
are themselves coupled by a second laser to electronically stable
molecules. For a degenerate system, the bosonic [fermionic] atoms
of mass $m_0$ [$m_+$] are represented by the field
$\psi_0(\br,t)$ [$\psi_+(\br,t)$], and the excited [stable]
fermionic molecules of mass $m_e=m_0+m_+$ [$m_-=m_e$] by the field
$\psi_e(\br,t)$ [$\psi_-(\br,t)$], with the boson (fermion) field operator
obeying commutation (anticommutation) relations. 
The Hamiltonian density for said non-ideal system is ${\cal H}
={\cal H}_0+{\cal H}_I$, where
\bml
\bea
\frac{{\cal H}_0}{\hbar} &=& -\Delta\psidag_-\psi_-
  +(\delta-\Delta)\psidag_e\psi_e
   +\lambda_{+-}\psidag_+\psidag_-\psi_-\psi_+
\nonumber\\
  &&+\sum_\sigma\psidag_\sigma
    \left[-\frac{\hbar\nabla^2}{m_\sigma}-\mu_\sigma
   +\lambda_{0\sigma}\psidag_0\psi_0\right]\psi_\sigma\,,
\label{H0}
\\
\frac{{\cal H}_I}{\hbar} &=& -\half\left[\left(\K_+\psidag_e\psi_+\psi_0
  +\Omega_-\psi_e\psidag_-\right) + \HC\right].
\eea
\label{FULL_HD}
\eml

The light-matter coupling due to
laser 1~(2) is ${\cal K}_+$ ($\Omega_-$), and the intermediate
(two-photon) laser detuning, basically the binding energy of the excited
(stable) molecular state relative to the photodissociation threshold, is
$\delta$ ($\Delta$). Particle trapping is implicit to the chemical
potential $\hbar\mu_\sigma$ ($\sigma=0,e,\pm$), and explicit traps can be
neglected for most practical purposes. Low-energy ($s$-wave) collisions
are accounted for by the boson-boson (boson-fermion, fermion-fermion)
interaction strength
$\lambda_{00}=2\pi\hbar a_{00}/m_0$
($\lambda_{0\pm}=4\pi\hbar a_{0\pm}/m_{0\pm}^*$,
$\lambda_{+-}=4\pi\hbar a_{+-}/m_{+-}^*$),
with $a_{\sigma_1\sigma_2}$ the $s$-wave scattering length
and $m_{\sigma_1\sigma_2}^*$ the reduced mass. Spontaneous decay,
included as $\Im[\delta]=-\Gamma$, is generally large enough
to justify the exclusion of excited-molecule collisions.

The backbone of free-bound-bound stimulated Raman
adiabatic passage is
{\em counterintuitive} pulse timing:
the two lasers are adjusted so that the level that
level with the most population is less strongly coupled to the
electronically-excited intermediate state, i.e., the
bound-bound pulse arrives first and the free-bound pulse arrives last, so
that the intermediate state population and the associated losses are kept
to a minimum. When the number of bosons is much greater than the number of
fermions,
$N_B\gg N_F$, the frequency scale for atom-molecule conversion is set by
$\Omega_+=\sqrt{\rho_B}\K_+$. Although a maximum
$N_B=100$ is used, qualitative scaling to large boson number is cinched
by assuming a density, $\rho_B=5\times 10^{14}\,{\rm cm}^{-3}$,
consistent with
$N_B=1.3\times 10^6$ Bose-condensed $^{85}$Rb atoms in a trap with radial
and axial frequencies
$\omega_r=100\times 2\pi\,$Hz and
$\omega_a=10\times 2\pi\,$Hz. For this density, a ballpark peak value for
the atom-molecule coupling is $\Omega_+\sim\Omega_0=2\pi\,$MHz.
A typical spontaneous decay rate is $\Gamma=10\times 2\pi\,$MHz.
The (assumedly negative) Fermi atom-molecule scattering length is
estimated as $|a_{+-}|=a_{00}=5.29\,$nm. The number of fermions is
restricted to $N_F=4$ for numerical ease, and large-particle-number
scaling is again ensured by picking a density,
$\rho_F=1.1\times 10^{12}\,\rm{cm}^{-3}$, consistent with $N_F=5\times
10^3$ $^{40}$K atoms in the same (mass-adjusted) trap as the bosons, so
that $\Lambda_{+-}=\lambda_{+-}/V=5.81\times 2\pi\,$Hz. Numerics are
further eased by restricting collisions between fermions to a bare
minimum. Also, $N_B\gg N_F$ means that collisions with condensate
atoms can be neglected. See Ref.~\cite{DAN03} for further details.

\begin{figure}
\centering
\includegraphics[width=8.0cm]{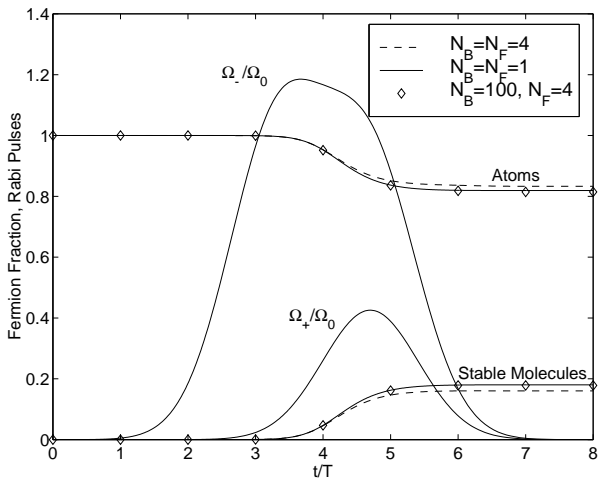}
\caption{
Creation of an arbitrary mixture of Fermi
degenerate atoms and molecules via fractional stimulated Raman adiabatic
passage.  Units of frequency are set by the choice
$\Omega_0=2\pi\,\rm{MHz}=1$, and
the pulse parameters are $\alpha=0.14\pi$, $T=5\times 10^3$, and
$\tau=0.7T$. For $N_B=100$ and $N_F=4$, fractional STIRAP exhibits no
visible difference from the $N_B=N_F=1$ case, while $N_B=N_F=4$
illustrates that many-body effects can limit the conversion
efficiency.}
\label{STIRAP}
\end{figure}

Of course, complete conversion would leave no
atoms to form Cooper pairs with molecules, so we pursue
fractional STIRAP~\cite{MAR91} via the Vitanov~\etal~\cite{MAR91}
pulseforms:
$\Omega_+(t)=\Omega_0\sin\alpha\exp[-\left(t-\tau\right)^2/T^2]$ and
$\Omega_-(t)=\Omega_0\exp[-\left(t+\tau\right)^2/T^2]
  +\Omega_0\cos\alpha\exp[-\left(t-\tau\right)^2/T^2]$, where
$\tan\alpha$ sets the final population fraction. Using Fock states and
a Hamiltonian derived from Eq.~\eq{FULL_HD}~\cite{MAC04}, we solve the
Schr\"odinger equation numerically. The key results are presented in
Fig.~\ref{STIRAP}. For
$N_F=4$ and
$N_B=100$, the system reproduces the single particle case ($N_B=N_F=1$),
i.e., the results are identical to those for a three-level atom, as
expected for a mostly-undepleted boson field. Atom-molecule collisions
are negligible for couplings as large as
$\Lambda_{+-}/\Omega_0=10^{-5}$, which is similar to a bosonic system,
lending confidence to the restricted-collision model.  Furthermore,
$N_F=N_B=4$ shows that many-body effects can limit molecular conversion.
This many-body effect is similar to the one-color BEC
case~\cite{JAV99}, and is not attributable to fermion statistics.

The important point is that STIRAP works basically as
expected, {\em even in the presence of collisions}. This
situation arises because, when $N_B\gg N_F$, the condensate density is
effectively fixed, and the associated mean-field contribution simply
amounts to a static bias that can be absorbed into the detuning; with
Fermi-Fermi collisions blocked, which is most likely for small
ground-state molecules, only collisions between the Fermi atoms and
molecules can move the system off the required two-photon resonance, and
STIRAP works better. In other words, we get the advantage of Bose
enhancement of the free-bound coupling ($\Omega_+\propto\sqrt{\rho_B}$),
without the mean-field shifts. While limited computational resources
preclude explicit investigation, these results should scale
qualitatively with increasing particle number, as for up to
$N_F=20$ in one-color production of Fermi molecules.

Now we are safe to presume the existence of an arbitrary admixture
of Fermi-degenerate atoms and molecules, and thus to consider any
subsequent anomalous pairing. Once the transient STIRAP pulses have
vanished, the system is described by ${\cal H}_0$ [Eq.~\eq{H0}] with
$\Delta=\delta=0$ and
$\sigma=0,\pm$. For equal-mass fermions, it is known that a fermion
density fluctuation gives rise to an effective chemical potential for the
bosons, which creates a bosonic density fluctuation, which in turn leads
to an effective chemical potential for the fermions. In other words,
phonons spawned by BEC density fluctuations are exchanged between the
fermions, altering their interaction. Just like lattice vibrations that
drive the attraction between degenerate electrons in superconductors, BEC
density fluctuations lead to an attractive interaction that can enhance
overall attractions, and thus Cooper pair formation~\cite{HEI00b}. 

Here the effective Fermi-Fermi scattering length is
\beq
\bar{a}_{+-} = a_{+-}
  \left[ 1+\frac{\ln(4e)^{2/3}}{\pi}\,k_Fa_{+-}
 -H\,\frac{\lambda_{0+}\lambda_{0-}}{\lambda_{00}\lambda_{+-}}\right],
\label{A_EFF}
\eeq
where
$H=\ln(1+x^2)/x^2$ with $x=\hbar k_F/m_0v_s$ and
$v_s=(\rho_B\hbar\lambda_{00}/m_0)^{1/2}$ is the speed of phonons
in BEC; $\hbar k_F\ll m_0v_s$ implies $H\approx 1$. In other words, the
effective scattering length can be written
$\bar{a}_{+-}/a_{+-}=1+\eta_{FF}-\eta_{FB}$, where $\eta_{FF}$
($\eta_{FB}$) is the contribution to atom-molecule interactions
from fermion-fermion (boson-fermion) fluctuations. Implicit to
expression~\eq{A_EFF} is the perturbative assumption
$\eta_{FB}\ll\eta_{FF}$. The immediate contrast with Ref.~\cite{HEI00b} is
that $\eta_{FB}<0$ is allowed.
For a weakly attractive system
($k_F|a_{\sigma_1\sigma_2}|,\rho_B|a_{\sigma_1\sigma_2}|^3\ll 1$), the
critical temperature for Cooper pairing is
\beq
T_c = 0.61\,T_F\,
  \exp\left[-\frac{\pi/4}{k_F|\bar{a}_{+-}|}\right],
\label{BCS_EQ}
\eeq
where
$T_F=\hbar(\mu_++\mu_-)R_M/k_B$ is the Fermi temperature with
$R_M=m_{+-}^*/\sqrt{m_+m_-}$. The Fermi wavevector, $k_F$, was taken as
the same for both species, so that $\mu_++\mu_-=(m_\pm/m_{+-}^*)\mu_\pm$
and $T_F=T_F^{(+)}\sqrt{m_+/m_-}\,$.

Continuing to focus on $^{87}$Rb-$^{40}$K,
$N_B=1.3\times 10^6$ BEC
atoms in a trap with $\omega_r=100\times 2\pi\,$Hz and
$\omega_a=10\times 2\pi\,$Hz yield a boson density 
$\rho_B=5\times 10^{14}\,\rm{cm}^{-3}$. A modest number of fermions,
$N_F=5\times 10^3$, means that the loss of condensate atoms in
molecule formation can be neglected, the condensate will absorb any heat
created by pairing-induced holes in the atomic Fermi sea~\cite{TIM01b},
and collapse instabilities~(Modugno \etal~\cite{TRU01}) are avoided.
Presuming that fractional STIRAP converts roughly 18\% of the initial
Fermi atoms into molecules (see Fig.~1), and that the fermions share
the same (mass-adjusted) trap, then
$\rho_\pm=1.1\times 10^{12}\,{\rm cm}^{-3}$, and the requirement of
equal Fermi wavenumbers for the atoms and molecules is met. For the
given parameters, the size of the BEC is
roughly an order of magnitude larger than the Fermi clouds, so that
overlap should not be an issue.

\begin{figure}
\centering
\includegraphics[width=8.0cm]{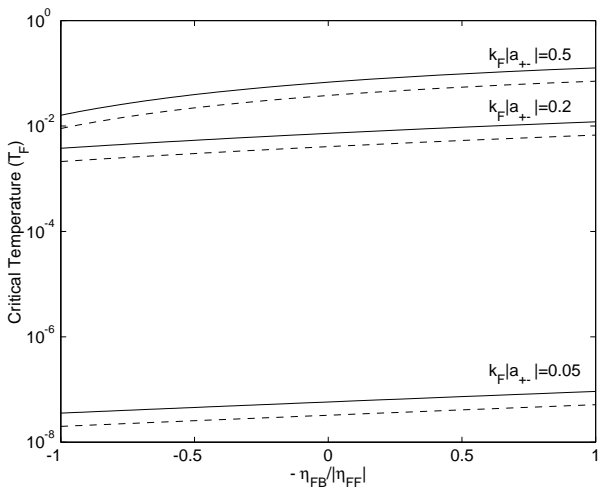}
\caption{Critical temperature for the superfluid transition to anomalous
atom-molecule pairs as a function of fermion-boson fluctuation
strength.  Calculations are for weak ($k_F|a_{+-}|=0.05$), marginally weak
(0.2), and marginally strong (0.5) interactions. The solid (dashed) curve
gives the critical temperature in units of the effective (atomic) Fermi
temperature.}
\label{CRIT_T}
\end{figure}

Figure~\ref{CRIT_T} summarizes our investigations. Under the above trap
conditions, and for $|a_{+-}|=a_{00}=5.29\,\rm{nm}$, we find the
weakness parameter $k_F|a_{+-}|=0.05$ and an unreasonably low
critical temperature. Nevertheless, if
the trap is modified to be anharmonic, then tighter confinement
ultimately means a diluteness parameter on the order of
$k_F|a_{+-}|=0.2$~\cite{SEA02}, and the situation is improved but still
out of reach of current technology
($T\sim0.05T_F$~\cite{REG03}). The only chance appears
to be for a tight trap and a large scattering length,
$k_F|a_{+-}|=0.5$; however, the theory is at best marginally
applicable in this regime and, besides, some other process (e.g.,
three-body recombination) would probably win out before superfluidity
could set-in to such a system.

\section{Crossover from a BEC of Trimers to atom-dimer Cooper pairs}
\label{TRIMERS}

While possible in principle, accessible atom-molecule superfluidity means
that next-generation technology must shed {\em another} order of
magnitude in temperature from the first generation~\cite{DEM99}, or that a
system with an attractive {\em and} a strong-but-not-too-strong
interaction will be found. Our opinion is that it will be worthwhile to
look for other ways of driving atom-molecule Cooper pairs, such as a
photoassociation or Feshbach resonance.

We therefore model an ideal degenerate
mixture of fermionic atoms and dimer molecules coupled by either a
Feshbach or photoassociation resonance to bound trimer molecules. An
ideal gas is chosen mainly because off-resonant particle-particle
interactions are generally too weak for practical purposes. The initial
fermionic atom-dimer state could be prepared using a
Raman scheme for photoassociating a degenerate Bose-Fermi mixture of
atoms (as in the previous section), and selectively removing the leftover
bosons. The atom-dimer $\leftrightarrow$ trimer resonance is expectedly
well resolved so that, once the initial atom-dimer state has been created,
transitions involving three free atoms are avoided. In contrast to the
all-boson case~\cite{MOO02}, ultracold transitions that involve a free
bosonic atom are Pauli blocked, i.e., the two identical fermionic
constituents of a given trimer may not form a bound state.

In second-quantization parley, a particle of mass $m_\sigma$ and momentum
$\hbar\bk$ is described by the creation operator $a_{\bk,\sigma}$. The
greek index corresponds mnemonically to the number of constituent atoms a
given particle contains: $3$ for bosonic trimers, 2 for fermionic dimers,
and 1 for fermionic atoms. All operators obey their (anti)commutation
relations. The microscopic Hamiltonian for such a freely-ideal system is
written:
%$H=H_0+H_I$,
\bea
\frac{H}{\hbar} &=&
\sum_{\bk,\sigma} \left[\left(\epsilon_{k,\sigma}-\mu_\sigma\right)
  \adag_{\bk,\sigma}a_{\bk,\sigma}\right]
\nonumber\\
&&-\frac{\K}{\sqrt{V}}
  \sum_{\bk,\bk'}\left(\adag_{\bk+\bk',3}a_{\bk,1}a_{\bk',2}+\HC\right).
%  +\adag_{\bk',2}\adag_{\bk,1}a_{\bk+\bk',3}\right).
%\nonumber\\
\label{MICRO_HAM}
\eea
The free-particle energy is defined by
$\hbar\epsilon_{k,\sigma}=\hbar^2 k^2/2m_\sigma$, and the chemical
potential by $\hbar\mu_\sigma$. In particular, the molecular
chemical potential is defined by
$\mu_3=2\mu-\delta_0$, where the bare detuning $\delta_0$ is a measure of
the binding energy of the trimer with
$\delta_0>0$ taken as above threshold. The (mode-independent)
atom-molecule coupling is $\K$, and
$V$ is the quantization volume.

The key realization is how to cast the Hamiltonian~\eq{MICRO_HAM} into a
readily diagonalized form. Consider a {\em time-dependent} unitary
transformation, which leaves the physics unchanged providing
$
H\rightarrow U^\dagger HU-iU^\dagger\partial_tU.
$
Given the generator
$U=\Pi_{\bk,\sigma}\exp[-it u_{\bk,\sigma}\adag_{\bk,\sigma}a_{\bk,\sigma}]$,
then $u_{\bk,3}=u_{\bk,1}+u_{\bk,2}$ implies
$[H,U]=0$ and, thus, $H\rightarrow H-iU^\dagger\partial_tU$.
Appropriately armed, apply the unitary transformation specified by
$u_{\bk,1(2)}=[\epsilon_{k,1(2)}-\epsilon_{k,2(1)}]/2$, which conveniently
corresponds to the special case
$u_{\bk,3}=0$, leaving the trimer term unchanged. The new
Hamiltonian reads:
\bea
\frac{H}{\hbar} &=&
\sum_\bk \left[\left(\epsilon_{k,3}+\delta_0-2\mu\right)\adag_{\bk,3}
a_{\bk,3}
  +(\varepsilon_k-\mu)\adag_{\bk,\sigma} a_{\bk,\sigma}\right]
\nonumber\\&&
-\frac{\K}{\sqrt{V}}\sum_{\bk,\bk'}\left(\adag_{\bk+\bk',3}
a_{\bk,1}a_{\bk',2}
  +\HC\right),
%  +\adag_{\bk',2}\adag_{\bk,1}b_{\bk+\bk',l}\right).
\label{HAM}
\eea
where the reduced free-particle energy is
$\hbar\varepsilon_k=\hbar^2 k^2/4m^*$, with $1/m^*=1/m_1+1/m_2$. 
Also, chemical equilibrium has been incorporated as $2\mu=\mu_1+\mu_2$. We
may now make a transformation to a dressed basis:
\bml
\beq
\left(\begin{array}{c}\alpha_{\bk,1}\\ \alphadag_{-\bk,2}\end{array}\right)
  =\left(\begin{array}{cc}
    \cos\theta_k & -e^{i\varphi}\sin\theta_k \\
    e^{-i\varphi}\sin\theta_k & \cos\theta_k
  \end{array}\right)\!\!
  \left(\begin{array}{c} a_{\bk,1}\\ \adag_{-\bk,2}\end{array}\right)\!\!,
\label{BOGOa}
\eeq
\beq
\alpha_{\bk,3} = a_{\bk,3} +\sqrt{V}\Phi\delta_{\bk,0},
\label{BOGOb}
\eeq
\label{BOGOT}
\eml
where $\delta_{\bk,0}$ is the Kronecker delta-function,
\bea
\frac{H}{\hbar} &=& \left(\delta_0 -2\mu\right)V|\Phi|^2 
  +\sum_\bk
    \left(\epsilon_{k,3}+\delta_0-2\mu\right)\alphadag_{\bk,3}\alpha_{\bk,3}
\nonumber\\&&
  +\sum_\bk \left[ \left( \varepsilon_k -\mu\right) 
    +\omega_k\left(\alphadag_{\bk,1}\alpha_{\bk,1}
      +\alphadag_{\bk,2}\alpha_{\bk,2}-1\right)\right].
\nonumber\\
\label{DIAG_HAM}
\eea
The condensate mean-field is
$\langle a_{0,3}\rangle/\sqrt{V}=e^{i\varphi}|\Phi|$, the mixing angle is
$\tan2\theta_k=|\Phi|\K/(\varepsilon_k-\mu)$, the quasiparticle frequency
is $\omega_k^2=(\varepsilon_k -\mu)^2 +|\Delta|^2$, and the gap
is $|\Delta|=\K|\Phi|$. The Hamiltonian~\eq{MICRO_HAM} is now lowest-order
diagonal.

Consider the mean-field Heisenberg equations
for the bosonic operator $a_{0,3}$ and the anomalous-pair-correlation
operator
$C_\bk={a}_{\bk,1}{a}_{-\bk,2}$ ({\it sans} chemical potential and
collective enhancement):
\bml
\bea
i\ddt\langle{a}_{0,3}\rangle&=&\delta_0\langle a_{0,3}\rangle
  -\frac{\K}{\sqrt{V}}\sum_{\bk} \langle C_\bk\rangle,
\label{HEIS_EQa}
\\
i\ddt\langle C_\bk\rangle
  &=&2\varepsilon_k\langle C_\bk\rangle
    -\frac{\K}{\sqrt{V}}\langle a_{0,3}\rangle.
\label{HEIS_EQb}
\eea
\label{HEIS_EQ}
\eml
Below threshold, simple Fourier analysis delivers the binding
energy, $\hbar\omega_B<0$, of the Bose-condensed trimers:
\beq
\omega_B-\delta_0-\Sigma(\omega_B)+i\eta=0,
\label{BIND}
\eeq
where $\Sigma(\omega_B)$ is the self-energy of the Bose molecules and
$\eta=0^+$. Incidentally, we show elsewhere~\cite{JAV04} that the real
poles of equation~\eq{BIND} fit the Regal \etal~\cite{REG03} data for
the binding energy of $^{40}\rm{K}_2$ molecules, and similar
measurements for a system of trimers would uniquely determine the
parameters of the present theory. On the other side of threshold,
the critical temperature for the transition to
effectively all superfluid atom-dimer pairs is derived from
Eq.~\eq{HEIS_EQa}:
\beq
T_c/T_F\simeq \exp\left(-\frac{\pi/4}{k_F|a_R|}\right).
\label{BCS}
\eeq
Here the resonant atom-dimer scattering length is
$a_R=-(4\pi m^*/\hbar)\K^2/\delta_0$. Also, we have taken a single Fermi
wavevector, $k_F$, for the atoms and the dimers, i.e.,
$\mu_1+\mu_2=\mu_{1(2)}m_{1(2)}/m^*$; assuming that the particles see the
same trap, adjusted for mass differences, equal Fermi wavevectors are
realized if the number of atoms and dimers satisfy
$N_2/N_1=(m_1/m_2)^{3/2}$. The effective Fermi temperature is
$k_B T_F/\hbar=(\mu_1+\mu_2)m^*/\sqrt{m_1m_2}$,
or $T_F=T_F^{(1)}\sqrt{m_1/m_2}$. Last but not least, it is easy to show
$|\Delta|\propto T_c$, so that
$|\Phi|\propto\exp\,(-\pi/4k_F|a_R|)$, and the trimer part of the
dressed BEC-pair admixture becomes larger near threshold
(increasing $|a_R|$), as expected.

Whereas the below-threshold regime of a trimer condensate is no doubt of
interest ($a_R>0$), both as a precursor to fermionic superfluidity and in
its own right, we keep our focus on attractive systems. The strongly
interacting regime is defined by
$k_F|a_R|\sim1$, indicating a transition to
predominantly atom-dimer Cooper pairs  at the critical temperature
$T_c\sim 0.45T_F$. Using
$m_3=m_1+2m_1$ as an example, which is akin to a system of
$^{6}\rm{Li}$ atoms and $^{7}\rm{Li}$-$^{6}\rm{Li}$ dimers, the required
dimer-atom fraction is $N_2/N_1=0.31$, the ratio of the effective
and atomic Fermi temperatures $T_F=0.7T_F^{(1)}$, and the critical
temperature
$T_c\sim 0.3T_F^{(1)}$. Although Eq.~\eq{BCS} is of dubious validity for
$T_c\lsimeq T_F$, it confirms that resonant association should in
principle drive superfluid pairing between atoms and dimer molecules at
transition temperatures within reach of present ultracold technology.

To rigorously identify the critical
temperature for the superfluid transition, it is necessary to go beyond
the effective atom-dimer theory, and explicitly include the bosonic
molecular state. Continuing to focus on a system of
$^{6}\rm{Li}$ atoms and $^{7}\rm{Li}$-$^{6}\rm{Li}$ dimers, we return to
the Hamiltonian~\eq{DIAG_HAM} and set $\epsilon_{k,3}\approx
\half\varepsilon_k$. We also introduce a
second molecular state, which can arise
because large detuning from one state brings the system into the
neighborhood of another bound state, or because of the presence of a
scattering resonance. The Hamiltonian~\eq{DIAG_HAM} is adapted simply by
making the substitution $\delta_0\rightarrow\delta_{0,l}$
($\Phi\rightarrow\Phi_l$, $\K\rightarrow\K_l$), and summing over the
index $l$; also, the gap becomes $|\Delta|=\K_1|\Phi_1|+\K_2|\Phi_2|$.
Here $\K_2\gsimeq\K_1$, and the system is tuned between the two levels so
that
$\delta_2>0$ and $\delta_1<0$.

\begin{figure}[t]
\centering
\includegraphics[width=8.0cm]{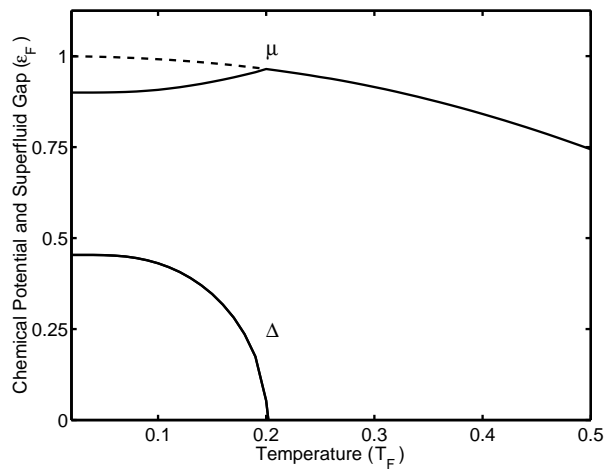}
\caption{Onset of the superfluid transition to a Bose-Einstein condensate
of trimers dressed by anomalous atom-dimer pairs. The non-zero
superfluid gap clearly lowers the system energy compared to the normal
state (dashed curve).
When the detuning is large and positive  ($\delta_2\gg\sqrt{\rho}\K_2$),
such as here, the system is mostly rogue pairs with a negligible fraction
of trimers.}
\label{SUPER}
\end{figure}

An algebraic system can be derived~\cite{FRI89} from the so-called grand
partition function,
$\Xi=\text{Tr}\exp\left( -\beta H\right)$, which we solve numerically for
the chemical potential as a function of temperature, which in turn should
display a characteristic cusp at the onset of superfluidity. Physically,
a cusp arises because the superfluid BEC-pair dressed state is lower in
energy than the normal state, implying the concurrent appearance of a
non-zero gap, as shown in Fig.~\ref{SUPER}. For positive detunings large
compared to the collective-enhanced coupling
($\delta_2\gg\sqrt{\rho}\K_2$ with
$\delta_1\approx-\delta_2$), the effective atom-dimer theory~\eq{BCS} with
$k_F|a_R|=1/2$ is an excellent working approximation. 
Also, the fraction of trimer is puny ($\sim 10^{-7}$), as per the large
detuning. Any $s$-wave collisional interactions are negligible
compared to the detuning, justifying the ideal-gas assumption.
The trap, albeit omitted, should actually favor the occurrence of
superfluid pairing~\cite{OHA02}.

Dimer molecules created near a Feshbach resonance are highly vibrationally
excited and, thus, characteristically long-range (K\"ohler
\etal~\cite{DON02}). Fermion-composite dimers are consequently
long-lived due to Pauli-suppressed vibrational
relaxation~\cite{PET03}, and there is no reason to expect otherwise
from Feshbach trimer states. In photoassociation, a two-color Raman
scheme is required to avoid spontaneous decay: a laser couples the atoms
to an electronically-excited intermediate trimer state, a second laser
couples the system to a ground-electronic trimer manifold, and the
intermediate trimer state is well-detuned. Long-range states are
available with photoassociation, although a Raman scheme also allows
access to stable lower-lying vibrational levels, which are much smaller
and less understood. Nevertheless, the molecular
fraction is negligibly small when the system is well above the
appropriate threshold, diminishing the chance for spontaneous
for decay of any kind.

\section{Summary and Outlook}
\label{OUT}

In conclusion, we highlight
that the term $\Omega=\sqrt{N}\,\kappa$ was previously
referred to as the Bose-enhanced free-bound coupling; however, this
behavior is now clearly independent of statistics, so that Bose
stimulation of free-bound association has nothing whatsoever to do with
Bose statistics, but is instead a many-body cooperative effect that
applies equally well to Fermi-degenerate systems. Consequently, we find
that collective Rabi oscillations and rapid adiabatic are feasible in
principle, and that the latter should also be possible in practice.
Similarly, stimulated Raman adiabatic passage from a degenerate Bose-Fermi
mixture to Fermi molecules should be possible, even in the presence of
collisions. On the other hand, while possible in principle, achieving
off-resonant atom-molecule superfluidity with current ultracold technology
means that a system with an attractive {\em and} a
strong-but-not-too-strong interaction must be found. Hence, we suggest
searching for an association resonance that combines atoms and diatomic
molecules into triatomic molecules, which leads to a crossover from a BEC
of trimers to atom-dimer Cooper pairs at a temperature reachable with
current techniques.

Each of the above results can be reinvestigated to determine the effect
of long-range dipolar interactions. Off hand, Rabi-like oscillations are
not likely to bear much fruit: the usual $s$-wave
collisions lead to mean-field shifts that make exact resonance, and thus
large-amplitude Rabi oscillations, difficult to achieve; the long-range
dipolar interaction will most probably make this difficult situation even
worse. But, we expect that rapid adiabatic passage survives the inclusion
of particle interactions, as in the all-boson case~\cite{ISH04}, and
the addition of a dipolar interactions for the molecules could sideline
this method for creating heteronuclear molecules. Moreover, STIRAP is
highly sensitive to particle-particle interactions~\cite{HOP01}, and it
is possible that inclusion of the long-range dipolar interaction could
moot this technique for anything other than simple heteronuclear systems
(e.g., $^6$Li-$^7$Li). Besides the formation of heteronuclear molecules
with an association resonance, the role of dipolar interactions in
atom-molecule Cooper pairing could also be investigated, which raises an
interest because the dipolar molecule will induce a dipole moment in the
atom (as opposed to pairing between two permanent {\em or} induced dipolar
particles~\cite{STO99}).

\begin{acknowledgements}
The authors kindly thank Robin C\^ot\'e for helpful hints.
We also gratefully acknowledge financial support from the Magnus Ehrnrooth
Foundation (O.D., J.P.), the Academy of Finland [K.-A.S. (project 206108)
and J.P. (project 204777)], as well as NSF and NASA [J.J. (PHY-0097974 and
NAG8-1428)].
\end{acknowledgements}

\end{document}